\documentclass[twocolumn,amsmath,amssymb,floatfix,groupedaddress,prl]{revtex4-1}

\usepackage{bm,graphicx,url,epsf,color}
\usepackage{amssymb}
\usepackage{amsthm}
\usepackage{cases}
\usepackage[colorlinks=true,linkcolor=blue,citecolor=blue]{hyperref}
\usepackage{comment}
\usepackage{mathrsfs}
\hypersetup{linkcolor=blue,urlcolor=blue,citecolor=blue}
\definecolor{purple}{rgb}{0.5,0,0.6}
\usepackage{ulem}
\definecolor{darkblue}{rgb}{0,0,0.5}
\definecolor{darkgreen}{rgb}{0,0.5,0}
\definecolor{darkred}{rgb}{.7,0,0}
\definecolor{purple}{rgb}{0.5,0,0.6}
\definecolor{orange}{rgb}{1,0.5,0}
\definecolor{grey}{rgb}{.6,.6,.6}
\definecolor{lightpink}{rgb}{1,0.7,0.75}
\definecolor{pink}{rgb}{1,0.4,0.58}
\definecolor{deeppink}{rgb}{1,0.08,0.58}

\renewcommand{\emph}[1]{\textit{#1}}

\usepackage{bbold}

\begin{document}
\title{Tunneling between two systems of interacting chiral fermions}
\author{D. B. Karki}
\affiliation{Materials Science Division, Argonne National Laboratory, Argonne, Illinois 60439, USA}
\author{K. A. Matveev}
\affiliation{Materials Science Division, Argonne National Laboratory, Argonne, Illinois 60439, USA}
\begin{abstract}
We develop a theory of tunneling between two systems of spinless chiral fermions. This setup can be realized at the edge of a quantum Hall bilayer structure. We find that the differential conductance of such a device in the absence of interactions has an infinitely sharp peak as a function of applied voltage. Interaction between fermions results in broadening of the conductance peak. We focus on the regime of strong interactions, in which the shape of the peak manifests well defined features associated with the elementary excitations of the system.
\end{abstract}
\maketitle
\emph{Introduction.---} Interactions between one-dimensional fermions give rise to a number of strongly correlated phenomena~\cite{Giamarchi}. Their complete understanding is generally beyond the scope of Landau Fermi liquid paradigm~\cite{Pine_Nozieres_book}. Instead, the low-energy properties of one-dimensional systems are typically described in terms of the Luttinger liquid theory~\cite{Haldane_1981}. The best known feature of this theory is the power-law scaling of the tunneling density of states at low energies~\cite{Kane_Fisher_1992,Furusaki_nagaosa_1993}. Luttinger liquid behavior has been experimentally observed in various one-dimensional systems, such as carbon nanotubes~\cite{CNT_1999, CNT_2004a, CNT_2004b} and semiconductor quantum wires~\cite{NW_2006}.

Similar phenomena have been predicted to emerge at the edges of fractional quantum Hall systems~\cite{Wen_1992}. In this case, the low energy excitations are described by the chiral Luttinger liquid theory, in which the properties of the system are controlled by the filling fraction $\nu$ of the corresponding quantum Hall state. For $\nu^{-1}$ being odd integer, the tunneling density of states of the chiral Luttinger liquid was predicted to follow the power-law behavior $D(\epsilon)\propto\epsilon^\chi$, where $\epsilon$ is the energy measured from the chemical potential and $\chi=\nu^{-1}-1$~\cite{Wen_1992}. These predictions have been confirmed experimentally~\cite{chiral_expt,Chang_RMP}. We note that although the very existence of fractional quantum Hall effect and, therefore, chiral Luttinger liquid at its edge, is due to the electron-electron interactions, the exponent in the above power law does not explicitly depend on the interaction strength.

In this work we consider the special case of integer quantum Hall effect with filling factor $\nu=1$. The edge state of this system can be modeled as a system of chiral one-dimensional fermions~\cite{Halperin_1982}. At $\nu=1$, the above power law gives a finite tunneling density of states at the chemical potential. Therefore, one may assume that the interactions do not significantly affect the properties of one-dimensional chiral fermions. On the other hand, recent work on spinless chiral fermions shows that the effects of interactions manifest themselves in the behavior of the spectral function of the system~\cite{KM1}. In particular, at fixed momentum $p$, the spectral function $A_p(\epsilon)$ has the shape of a peak with the width proportional to the interaction strength.

The main difference between the density of states $D(\epsilon)$ and the spectral function $A_p(\epsilon)$ is that the latter describes the response of the system when a particle is added to or removed from it has a fixed momentum $p$. In one dimension, such physical processes can be probed experimentally in devices with momentum-conserved tunneling setup~\cite{Kang_2000, Auslaender_2002, Yacobi_2007}. In recent experiments~\cite{Patlatiuk_2018, Patlatiuk_2020}, momentum-conserved tunneling between a quantum wire and a quantum Hall system was used to probe the edge states with unprecedented accuracy. Alternatively, momentum-conserved tunneling can be achieved in quantum Hall bilayer devices~\cite{Eisenstein_1992,Eisenstein_2011, Klitzing_2012,Eisenstein_2013} operated in the regime, where transport between the layers is dominated by edge states rather than the gapped excitations in the bulk. Given these experimental advances, it is important to study theoretically how interactions affect momentum-conserved tunneling in systems of interacting chiral fermions.

In this paper, we develop a general theory of momentum-conserved tunneling between two systems of spinless one-dimensional chiral fermions. We account for interactions both within and between the two systems and focus on the limit of strong interactions. As an application of our theory, we study tunneling between the edge states in a quantum Hall bilayer device. We show that the differential conductance of such a device has a sharp peak as a function of applied voltage. We find that the shape of the peak shows well defined features associated with the elementary excitations of the system. 

\begin{figure}[b]
\begin{center}
\includegraphics[scale=0.62]{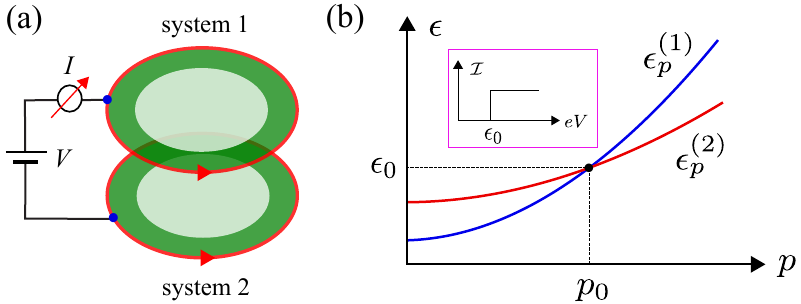}
\caption{(a) The schematic representation of the setup under consideration, which can be realized in quantum Hall bilayer devices. (b) The sketch of the free fermion dispersions $\epsilon^{(j)}_p$. In the absence of interactions, current shows step-like behavior as illustrated in the inset.}\label{fig1}
\end{center}
\end{figure}

\emph{Model.---} 
We consider two systems of interacting one-dimensional spinless chiral fermions separated by a short distance, as illustrated in Fig.~\ref{fig1}(a).  This setup is described by the Hamiltonian $\hat{H}=\hat{H}_0+\hat{H}_T$, with 
\begin{align}
\hat{H}_0=&\sum_p\epsilon^{(1)}_{ p} c_{1,p}^\dagger c_{1,p}+\sum_p\epsilon^{(2)}_{ p} c_{2,p}^\dagger c_{2,p}\nonumber\\
&+\frac{1}{L}\sum_{j=1}^2\sum_{pp'\atop q>0}V^{(j)}(q) c^\dagger_{j, p+q}c_{j, p}c^\dagger_{j, p'}c_{j, p'+q}\nonumber\\
&+\frac{1}{L}\sum_{pp'\atop q>0}\tilde{V}(q)\Big( c^\dagger_{1, p+q}c_{1, p}c^\dagger_{2, p'}c_{2, p'+q}\nonumber\\
&\;\;\;\;\;\;\;\;\;\;\;\;\;\;\;\;\;\;\;\;\;\;\;+c^\dagger_{2, p+q}c_{2, p}c^\dagger_{1, p'}c_{1, p'+q}\Big).\label{mod2}
\end{align}
Here the operator $c_{j, p}$ annihilates a fermion in the state with momentum $p$ in system $j$, with $j=1,2$. The energy of fermions in system $j$ is represented by $\epsilon^{(j)}_p$, which is assumed to be a monotonic function of $p$. Motivated by the experiments on quantum Hall bilayer devices, we assume that the two systems of chiral fermions are of the same length $L$ and have similar properties, but are not identical. The interactions within each system and between them are expressed in terms of the Fourier transformed interaction potentials $V^{(j)}(q)$ and $\tilde{V}(q)$, respectively. In this paper, we assume that these quantities are positive, which corresponds to repulsive interactions. The operator
\begin{equation}
\hat{H}_T=\sum_{p}\left(\gamma_{p}c^\dagger_{1, p}c_{2 p}+\gamma^*_{p}c^\dagger_{2, p}c_{1 p}\right),\label{mod4}
\end{equation}
describes momentum-conserved tunneling between the two systems; $\gamma_p$ stands for the tunneling matrix element.

To calculate the tunneling current, we start with the operator $\hat{I}=\left(ie/\hbar\right)\sum_p\left(\gamma_p c^\dagger_{1,p}c_{2,p}-{\rm h.c.}\right)$ accounting for the charge current from system 1 to system 2. Assuming weak tunneling regime, we proceed with the evaluation of charge current $\left<\hat{I}\right>$ perturbatively in $\hat{H}_T$ and, retaining only the terms of the lowest order in $\gamma_p$. The resulting expression for the current per unit length $\mathcal{I}=\left<\hat{I}\right>/L$ takes the form
\begin{subequations}
\begin{equation}
\mathcal{I}=\frac{e}{L\hbar^2}\sum_p|\gamma_p|^2\int^\infty_{-\infty}dt\left[\mathcal{F}^{(12)}_p(t)
-\mathcal{F}^{(21)}_p(-t)\right],\label{mod5a}
\end{equation}
\begin{equation}
\mathcal{F}^{(jj')}_p(t)=\Big<c^\dagger_{j, p}(t)c_{j', p}(t)c^\dagger_{j', p}(0)c_{j, p}(0)\Big>.\label{mod5b}
\end{equation}
\end{subequations}

In the absence of interactions, the two particle average~\eqref{mod5b} decouples into the product of single particle averages, which can be calculated straightforwardly. In this case, Eq.~\eqref{mod5a} gives the expression for current per unit length at zero temperature in the form
\begin{equation}
\mathcal{I}=\frac{e}{\hbar^2}\sum_m|\gamma_{p_m}|^2\frac{\left[\theta\left(\epsilon_{m}-\mu_2\right)-\theta\left(\epsilon_{m}-\mu_1\right)\right]}{\left|\left.\partial_p(\epsilon^{(1)}_p-\epsilon^{(2)}_p)\right|_{p=p_m}\right|},\label{mod6a}
\end{equation}
where $\theta(x)$ is the unit step function and $\mu_j$ stands for the chemical potential of the system $j$. Momentum-conserved tunneling is possible when the energies of fermions in two systems become equal at a certain momentum, i.e., $\epsilon^{(1)}_{p}=\epsilon^{(2)}_{p}$. In general, there are several solutions $p=p_m$ of this equation; the corresponding energies are denoted by $\epsilon_m$, where $m=0, 1,\ldots$. Figure~\ref{fig1}(b) illustrates the case of a single solution with $p=p_0$.

To study the current-voltage characteristic of the device, we introduce voltage as $eV=\mu_1-\mu_2$ and also assume that the chemical potential $\mu_2$ is fixed and chosen as the origin of energy, i.e., $\mu_2=0$. The current~\eqref{mod6a} shows steps as a function of voltage. Current-voltage characteristic in the case of a single solution with $\epsilon^{(1)}_{p}=\epsilon^{(2)}_{p}=\epsilon_0>0$ is shown in the inset of Fig.~\ref{fig1}(b). The corresponding differential conductance per unit length $G=\partial\mathcal{I}/\partial V$ has an infinitely sharp peak at $eV=\epsilon_0$.

The main objective of this work is to investigate how the presence of strong interactions within and between the systems affects the current-voltage characteristic shown in the inset of Fig.~\ref{fig1}(b). In this case, the average~\eqref{mod5b} no longer decouples. We focus on the low energy regime, where calculations in the presence of strong interactions are simplified greatly upon the bosonization of the Hamiltonian of the system.

\emph{Bosonization.---} We follow the standard bosonization procedure~\cite{Haldane_1981} to reformulate the problem in bosonic variables using the transformation
\begin{equation}
\psi_j(x)=\frac{\hat{u}_j}{\sqrt{L}}e^{i p^{(j)}_F x/\hbar} e^{i\varphi^\dagger_j(x)}e^{i\varphi_j(x)}.\label{meth1}
\end{equation}
Here $\psi_j(x)$ annihilates a fermion at position $x$ in system $j$; it is related to the operator $c_{j, p}$ by the Fourier transformation. The operator $\hat{u}_j$ lowers the particle number $N_j$ in system $j$ by $1$, i.e., $[\hat{u}_j, \hat{N}_j]=\hat{u}_j$, and also includes the Klein factor emerging in the bosonization procedure. The Fermi momentum of system $j$ is denoted by $p^{(j)}_F$.
The bosonic fields $\varphi_j(x)$ are defined by
\begin{equation}
\varphi_j(x)=-i\sum_{l=1}^\infty\frac{1}{\sqrt{l}}e^{iq_lx/\hbar}b_{j, l},\;\;q_l=\frac{2\pi\hbar}{L}l,\label{meth2}
\end{equation}
with $b_{j, l}$ being the bosonic annihilation operators.

To express $\hat{H}_0$ in terms of bosonic variables, we first note that in the strong interaction regime, the curvature of electronic dispersion can be neglected~\cite{KM1}. Thus, we linearize the fermion dispersion $\epsilon^{(j)}_p$ near the Fermi point,
\begin{equation}
\epsilon^{(j)}_p=\mu_j+ v^{(j)}_F\left(p-p^{(j)}_F\right),\label{aammaa}
\end{equation}
where the corresponding Fermi velocity is denoted by $v^{(j)}_F$. We note that the Fermi momentum $p^{(j)}_F$ depends on the chemical potential $\mu_j$ and that this dependence is affected by the interactions~\cite{SI}. Using Eqs.~\eqref{meth1}$-$\eqref{aammaa}, we express the operator $\hat{H}_0$ in terms of bosonic variables as
\begin{subequations}
\begin{align}
\hat{H}_0 =&\sum_{l=1}^\infty\left(\varepsilon^{(1)}_{ l}b^\dagger_{1, l}b_{1, l}+\varepsilon^{(2)}_{ l}b^\dagger_{2, l}b_{2, l}\right)+\mu_1\hat{N}_1+\mu_2\hat{N}_2\nonumber\\
&\;\;\;\;\;\;\;\;\;\;\;\;\;\;\;\;+\sum_{l=1}^\infty U_l\left(b^\dagger_{1, l}b_{2, l}+b^\dagger_{2, l}b_{1, l}\right),\label{meth3}
\end{align}
\begin{equation}
\varepsilon^{(j)}_{ l}=\left[v^{(j)}_F+\frac{V^{(j)}(q_l)}{2\pi\hbar}\right]q_l,\quad U_l=\frac{\tilde{V}(q_l)}{2\pi\hbar}q_l.\label{meth4}
\end{equation}
\end{subequations}
It is convenient to bring Eq.~\eqref{meth3} to the diagonal form
\begin{equation}
\hat{H}_0=\sum_{l=1}^\infty\!\left(\tilde{\varepsilon}^{(1)}_l \tilde{b}^\dagger_{1, l}\tilde{b}_{1, l}{+}\tilde{\varepsilon}^{(2)}_l \tilde{b}^\dagger_{2, l}\tilde{b}_{2, l}\right)+\mu_1\hat{N}_1+\mu_2\hat{N}_2\label{meth6}
\end{equation}
by performing the transformation
\begin{equation}
\begin{pmatrix}
b_{1, l}\\
b_{2, l}
\end{pmatrix}=\begin{pmatrix}
\cos \theta_l & {-}\sin \theta_l\\
\sin \theta_l & \cos \theta_l
\end{pmatrix}\begin{pmatrix}
\tilde{b}_{1, l}\\
\tilde{b}_{2, l}
\end{pmatrix}\!,\;\tan 2\theta_l=\frac{2 U_l}{\varepsilon^{(1)}_{ l}{-}\varepsilon^{(2)}_{ l}}.\label{meth5}
\end{equation}
The energies $\tilde{\varepsilon}^{(i)}_{ l}$ in Eq.~\eqref{meth6} are defined by
\begin{equation}
\tilde{\varepsilon}^{(j)}_{ l} =\frac{\varepsilon^{(1)}_{ l}{+}\varepsilon^{(2)}_{ l}}{2}-\frac{(-1)^j}{2}\sqrt{\left(\varepsilon^{(1)}_{ l}-\varepsilon^{(2)}_{ l}\right)^2+4 U_l^2},\label{meth7}
\end{equation}
where we assumed $\varepsilon^{(1)}_{ l}>\varepsilon^{(2)}_{ l}$~\footnote{The interaction parameters $V^{(j)}$ and $V^{(12)}$ are independent. In general, the interactions within the systems are stronger than those between the systems. Therefore, we proceed with the assumption $V^{(j)}>V^{(12)}$, which guarantees $\tilde{\varepsilon}_l^{(1)}>\tilde{\varepsilon}_l^{(2)}>0$. }.


\emph{Expression for the charge current.---} 
To evaluate the charge current~\eqref{mod5a}, we need an expression for the time dependent operator $\psi_j(x, t)$. In addition to the time dependence of the bosonic field $\varphi_j$ in Eq.~\eqref{meth1}, the Hamiltonian~\eqref{meth6} generates the time dependence of the lowering operator $\hat{u}_j(t)=\hat{u}_je^{-i\mu_j t/\hbar}$. Therefore,
\begin{equation}
\psi_j(x, t)=\frac{\hat{u}_j }{\sqrt{L}}e^{-i\mu_j t/\hbar}e^{i p^{(j)}_F x/\hbar} e^{i\varphi^\dagger_j(x, t)}e^{i\varphi_j(x, t)}.\label{curr3}
\end{equation}
Using Eqs.~\eqref{mod5a} and~\eqref{curr3}, we proceed with the evaluation of charge current. In the thermodynamic limit $L\to\infty$, we find the following expression for the current per unit length at zero temperature~\cite{SI}
\begin{align}
&\mathcal{I}=\frac{ e|\gamma|^2}{\hbar^2 }\int^{\infty}_{-\infty} dq\int^{\infty}_{-\infty}d\varepsilon\;\tilde{A}^{(1)}\left(q, \varepsilon\right)\nonumber\\
&\;\;\;\;\;\;\;\;\;\;\;\;\;\;\;\;\;\;\;\;\times\tilde{A}^{(2)}\left(\Delta p_F-q, \Delta\mu-\varepsilon\right).\label{current4}
\end{align}
Here $\Delta p_{ F}=p^{(1)}_{ F}-p^{(2)}_{F}$ and $\Delta\mu=\mu_1-\mu_2>0$. At low energies, tunneling of fermions is confined to the vicinity of the Fermi level and thus to arrive at Eq.~\eqref{current4} we neglected the momentum dependence of tunneling matrix element, $\gamma_p\to\gamma$. The function $\tilde{A}^{(j)}$ is defined by
\begin{align}
\tilde{A}^{(j)}(q, \varepsilon) &=\lim_{L\to\infty}\int^\infty_{-\infty} \frac{dt}{2\pi\hbar} \int^L_{0}\frac{dy}{L}\; e^{-i\left(q y-\varepsilon t\right)/\hbar}\nonumber\\
&\;\;\;\;\;\;\times \exp\left(\sum_{l=1}^\infty\frac{1}{l}\; e^{i\left[q_l y-\tilde{\varepsilon}^{(j)}(q_l)t\right]/\hbar}\right).\label{spfn}
\end{align}
We note that, in the absence of interactions between the two systems, $\tilde{A}^{(j)}$ represents the spectral function of the system $j$~\cite{KM1}. In this case, the charge current given by Eq.~\eqref{current4} is proportional to the convolution of the spectral functions of the two systems.

The form of the charge current given by Eq.~\eqref{current4} reflects the momentum and energy conservation in our system. To demonstrate this, we consider a process of tunneling of fermion from system 1 to system 2. This particle exchange process produces the overall change in the energy of the system $\varepsilon_1+ \varepsilon_2-\Delta \mu$ as seen from Eq.~\eqref{meth6}. Here, $ \varepsilon_j$ is the total energy of bosonic excitations in the branch $j$ after tunneling. Similarly, the change in momentum is $q_1+ q_2-\Delta p_F$, with $q_j$ being the total momentum of the bosonic branch $j$. Therefore, the conservation of energy and momentum requires $ \varepsilon_2=\Delta\mu- \varepsilon_1$ and $ q_2=\Delta p_F- q_1$, which is reflected in Eq.~\eqref{current4}. To study the current~\eqref{current4} in more detail, we need to specify a particular form of the interactions.

\emph{Contact interactions.---} In the limit of extremely short range interactions, the energy of bosonic excitations given in Eq.~\eqref{meth7} is linear in momentum, i.e., $\tilde{\varepsilon}^{(j)}(q)=\tilde{v}_jq$, where
\begin{subequations}\label{contact1}
\begin{equation}
\tilde{v}_{1,2} =\frac{v_1+v_2\pm\sqrt{(v_1-v_2)^2+4 v_{12}^2}}{2},
\end{equation}
\begin{equation}
v_j=v^{(j)}_F+\frac{V^{(j)}(0)}{2\pi\hbar}, \quad v_{12}=\frac{\tilde{V}(0)}{2\pi\hbar}.
\end{equation}
\end{subequations}
To study the charge current~\eqref{current4}, we also need an expression for the voltage dependence of $\Delta p_F$. Setting again $\mu_1=eV$ and $\mu_2=0$, we obtain~\cite{SI}
\begin{equation}
\Delta p_F(V)=\Delta p_F(0)+\frac{eV}{ \overline{v}},\quad\overline{v}=\frac{v_1v_2-v^2_{12}}{v_2+v_{12}}.\label{contact2}
\end{equation}

For the linear dispersion $\tilde{\varepsilon}^{(j)}(q)=\tilde{v}_jq$, Eq.~\eqref{spfn} yields $\tilde{A}^{(j)}(q, \varepsilon)=\theta(\varepsilon)\delta(\varepsilon-\tilde{v}_j q)$. Using Eq.~\eqref{current4}, we then find 
\begin{equation}
\mathcal{I} =\frac{e|\gamma|^2}{\hbar^2\left(\tilde{v}_1{-}\tilde{v}_2\right)}\left[\theta\left(eV{-}eV_1\right)+
\theta\left(eV{-}eV_2\right){-}1\right],\label{contact3}
\end{equation}
where $V_j$ is found from the condition $eV_j=\tilde{v}_j\Delta p_F(V_j)$, and is given by $eV_j=\tilde{v}_j\overline{v}\Delta p_F(0)/\left(\overline{v}-\tilde{v}_j \right)$. Although Eq.~\eqref{current4} was written for $\Delta\mu>0$, the result~\eqref{contact3} applies for both positive and negative voltage $V$.

In general, the charge current~\eqref{contact3} has two steps positioned at the voltages $V_j$, with $V_1$ and $V_2$ being of the opposite sign due the inequality $\tilde{v}_1\geq\overline{v}\geq\tilde{v}_2$. At small $\Delta p_F(0)$, voltages $V_1$ and $V_{2}$ are small and thus we are in the regime of applicability of bosonization theory. The number of steps reduces to one in two special cases. First, in the absence of interactions between the two systems, i.e., $\tilde{V}(0)=0$ and thus $v_{12}=0$, we have $V_1\to\infty$. Second, in the limit of $v_1=v_2$, we get $V_2\to\infty$.

It is instructive to compare this behavior with that of non-interacting systems, where current is given by Eq.~\eqref{mod6a}. Assuming linear dispersions in Eq.~\eqref{mod6a}, we find only one step. The absence of the second step is due to the fact that the applied voltage changes only $p_F^{(1)}$, while $p_F^{(2)}$ is fixed. Different application of voltage, such as $\mu_{1, 2}=\pm eV/2$, would result in two steps in $\mathcal{I}(V)$. In the special case, when the Fermi velocities of the two systems are identical, current~\eqref{mod6a} vanishes. This is in contrast to the case of interacting systems, where even at $v_1=v_2$, current~\eqref{contact3} yields a step at voltage $V_1$. 

\begin{figure}[t]
\includegraphics[scale=0.38]{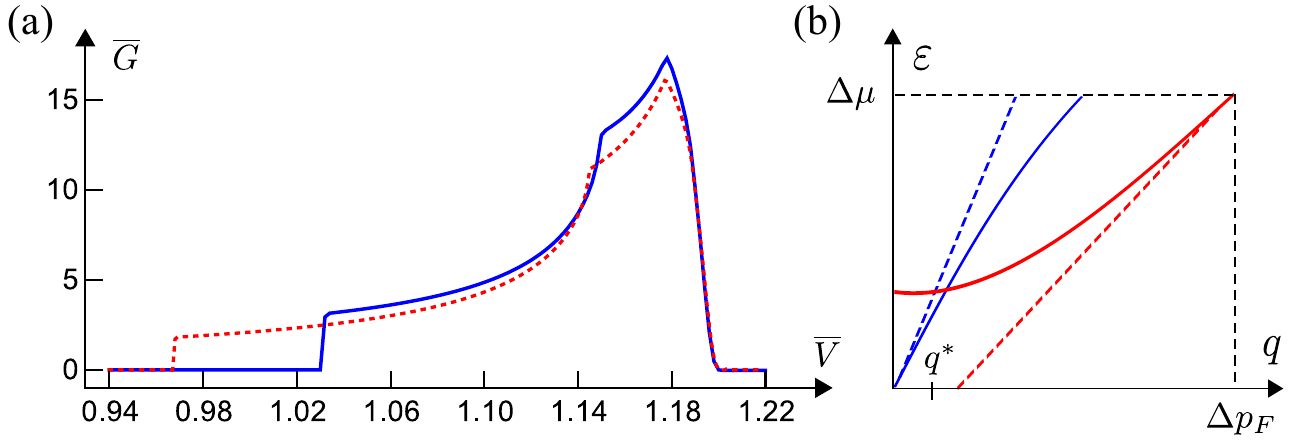}
\caption{(a) The dimensionless form of differential conductance defined by $\overline{G}=\partial\overline{\mathcal{I}}/\partial\overline{V}$, where $\overline{\mathcal{I}}=\left(\hbar^2 v_1/ e |\gamma|^2\right) \mathcal{I}$ and $\overline{V}=eV/\left[\Delta p_F(0)v_1\right]$. The solid blue curve is obtained from Eq.~\eqref{current4} for the choice of parameters $v_2=0.5 v_1$, $v_{12}=0.3 v_1$, $\eta_2=0.8\eta_1$, $\eta_{12}=0.6\eta_1$, and $\eta_1[\Delta p_F(0)]^2/(2\pi\hbar v_1)=0.005$. The dotted red curve shows the conductance to leading order in small non-linearity given by Eq.~\eqref{short9} for the same values of parameters as that of the blue curve. (b) The blue and red lines are the plots of the spectra $\tilde{\varepsilon}_1(q)$ and $\Delta\mu-\tilde{\varepsilon}_2(\Delta p_F-q)$. Dashed lines show the linear asymptotes of corresponding bosonic spectra. Only the small traingular region with typical momentum $q^*$ contributes to the current~\eqref{current4}.}\label{fig2}
\end{figure}

\emph{Short range interactions.---} When the range of interactions is non-zero, the non-linearity of bosonic spectrum should be taken into account. Assuming the interactions are of short range, we substitute $\varepsilon^{(j)}(q)=v_j q-\eta_j q^3/\left(2\pi\hbar\right)$ and $U(q)=v_{12} q-\eta_{12} q^3/\left(2\pi\hbar\right)$ into Eq.~\eqref{meth7} and obtain the bosonic spectra $\tilde{\varepsilon}^{(j)}$ in the form 
\begin{equation}
\tilde{\varepsilon}^{(j)}(q)=\tilde{v}_j q-\frac{\tilde{\eta}_j}{2\pi\hbar}q^3. \label{short1}
\end{equation}
Here the parameter $\tilde{\eta}_j>0$ accounts for the curvature of the bosonic spectrum; its expression in terms of microscopic parameters of the model is given in Ref.~\cite{SI}. For the spectrum~\eqref{short1}, the function $\tilde{A}^{(j)}$ takes the form~\cite{KM1}
\begin{align}
\tilde{A}^{(j)}(q, \varepsilon)=\frac{2\pi\hbar}{\tilde{\eta}_j q^3}\;a\left(\frac{2\pi\hbar}{\tilde{\eta}_j q^3}\left[\varepsilon-\tilde{\varepsilon}^{(j)}(q)\right]\right).\label{short4}
\end{align}
Here $a(x)$ is a universal function, which vanishes outside the region $0<x<1$ and has singularities at $x=1-n^{-2}$, where $n=1, 2,\ldots$. The position of the $n$th singularity corresponds to the energy $\varepsilon=n\tilde{\varepsilon}^{(j)}(q/n)$ and thus reflects the nature of the many-body excitations in the system.

Substitution of Eq.~\eqref{short4} into Eq.~\eqref{current4} allows one to study the differential conductance per unit length $G=\partial \mathcal{I}/\partial V$ for $\Delta\mu=eV>0$. Thus, assuming $\Delta p_F(0)>0$, we now study the differential conductance in the vicinity of the step at $V=V_2$. For a particular choice of parameters, the voltage dependence $G(V)$ obtained numerically from Eq.~\eqref{current4} is plotted in Fig.~\ref{fig2}(a). The $\delta$-function peak of conductance in the case of contact interactions is now broadened because of the non-linearity of the bosonic spectrum. The shape of the peaks shows clear singularities at certain values of voltage.

To understand the shape of the peak of differential conductance shown in Fig.~\ref{fig2}(a), we study the current~\eqref{current4} in leading order in small parameters $\tilde{\eta}_j$. To this end, we first note that the function $\tilde{A}^{(j)}(q, \varepsilon)$ defined by Eq.~\eqref{spfn} is finite only when both momentum and energy are positive. This constraint restricts the range of integration in Eq.~\eqref{current4} to $0<q<\Delta p_F$ and $0<\varepsilon<\Delta\mu$, which corresponds to the rectangle shown in Fig.~\ref{fig2}(b). In addition, the concavity of spectrum~\eqref{short1} guarantees that the total energy $\varepsilon_j$ of any set of bosonic excitations with the total momentum $q$ is in the range $\tilde{\varepsilon}^{(j)}(q)\leq \varepsilon_j< \tilde{v}_jq$. (This is the reason why $a(x)$ in Eq.~\eqref{short4} vanishes outside the region $0<x<1$.) It is then clear that only the region defined by $\tilde{\varepsilon}^{(1)}(q)\leq \varepsilon< \tilde{v}_1q$ and $\Delta\mu-\tilde{\varepsilon}^{(2)}(\Delta p_F-q)\geq \varepsilon> \Delta\mu-\tilde{v}_2(\Delta p_F-q)$ contributes to the current~\eqref{current4}. This region can be visualized by plotting the spectra $\tilde{\varepsilon}^{(1)}(q)$ and $\Delta\mu-\tilde{\varepsilon}^{(2)}(\Delta p_F-q)$ along with their linear asymptotes as illustrated in Fig.~\ref{fig2}(b), where we took $\Delta\mu\sim eV_2$, i.e., the red dotted line passes near the point $(q, \varepsilon)=(0, 0)$. From the figure, it is seen that only the small triangular region formed by crossing of the spectrum of the slow branch (red) with that of the fast branch (blue) and its linear asymptote (dashed blue) contributes to the current~\eqref{current4}.

In the case of weak non-linearity, i.e., $(\Delta p_F)^2\tilde{\eta}_j\ll 2\pi\hbar\tilde{v}_j$, typical momentum $q^*$ within this triangular region is small, $q^*\ll\Delta p_F$. Thus the broadening of $\tilde{A}^{(1)}(q, \varepsilon)$ due to the non-linearity of spectrum can be neglected compared to that of $\tilde{A}^{(2)}(\Delta p_F-q, \Delta\mu-\varepsilon)$, i.e., we can replace $\tilde{A}^{(1)}(q, \varepsilon)\to \theta(\varepsilon)\delta(\varepsilon-\tilde{v}_1 q)$. Equation~\eqref{current4} then reduces to a single integral. The result to leading order in $\tilde{\eta}_2$ takes the compact form~\footnote{A better agreement with the numerical result shown in Fig.~\ref{fig2}(a) is obtained when higher order corrections are taken into account~\cite{SI}.}
\begin{subequations}\label{short9}
\begin{equation}
\mathcal{I} =\frac{e|\gamma|^2}{\hbar^2(\tilde{v}_1{-}\tilde{v}_2)}\int^{1+\lambda}_{0}\!\!\!dx\;a(x),
\end{equation}
\begin{equation}
\lambda=\frac{2\pi\hbar e \left(V-V_2\right)}{ \tilde{\eta}_2\left[\Delta p_F(0)\right]^3}\left(1-\frac{\tilde{v}_2}{\overline{v}}\right)^4.
\end{equation}
\end{subequations}

Current~\eqref{short9} vanishes at voltages corresponding to $\lambda<-1$, then grows monotonically and reaches a plateau at voltage $V_2$, where $\lambda$ vanishes. Differentiation of Eq.~\eqref{short9} with respect to voltage yields conductance peak with the shape essentially identical to that of the universal function $a(x)$. As shown in the Fig.~\ref{fig2}(a), the differential conductance obtained from Eq.~\eqref{short9} reproduces all the features of that obtained from Eq.~\eqref{current4} numerically. 

\emph{Discussion.---} Our results can be tested in experiments with quantum Hall bilayer structures. The sizes of the devices in the existing experiments~\cite{Eisenstein_1992,Eisenstein_2011, Klitzing_2012,Eisenstein_2013} are large, and as a result the transport is dominated by the gapped excitations in the bulk. By using smaller size structures, transport properties dominated by the edge states can be investigated. Typically, in quantum Hall devices, the electron density is controlled by nearby gates, which screen the Coulomb interactions and thus the case of short range interactions discussed here will be relevant.



To summarize, we presented a low-energy theory of momentum-conserved tunneling between two systems of spinless one-dimensional chiral fermions in the regime of strong interactions. We applied our theory to study transport in small quantum Hall bilayer devices. We showed that the differential conductance of such a device has a distinctive shape with nontrivial features reflecting the many-body physics of interacting chiral fermions.  

This work was supported by the U. S. Department of Energy, Office of Science, Basic Energy Sciences, Materials Sciences and Engineering Division.
%
\vfill
\pagebreak
\widetext
\begin{center}
\textbf{\large \large{SUPPLEMENTAL MATERIAL}}
\end{center}

\setcounter{equation}{0}
\setcounter{figure}{0}
\setcounter{table}{0}
\makeatletter
\renewcommand{\theequation}{S\arabic{equation}}
\renewcommand{\thefigure}{S\arabic{figure}}
\renewcommand{\bibnumfmt}[1]{[S#1]}
\renewcommand{\citenumfont}[1]{S#1}

In this supplemental material, we outline the derivation for the main equations used in the main text.
\section{General expression for the charge current}
In this section we outline the derivation for the expression of charge current per unit length, which is quoted in Eq.~(13) of the main text. Since $\psi_j(x, t)$ defined in Eq.~(12) and the fermion operator $c_{j, p}$ introduced in Eq.~(1) are related by the Fourier transformation, we first express the average current per unit length given by Eq.~(3) in terms of $\psi_j(x, t)$. The result in the thermodynamic limit can be expressed into the form
\begin{align}
\mathcal{I}
&=\lim_{L\to\infty}\frac{e|\gamma|^2}{ L^4\hbar^2}\int^\infty_{-\infty}dt\; e^{i\Delta\mu t/\hbar}\;\int^L_0 dx_1dx_2dx_3 \Bigg[e^{-i(p^{(1)}_F-p^{(2)}_F)(x_2-x_3)/\hbar} \mathcal{Q}_1(x_1, x_2, x_3, x_1-x_2+x_3, t)\nonumber\\
&\;\;\;\;\;\;\;\;\;\;\;\;\;\;\;\;\;\;\;\;\;\;\;\;\;\;\;\;\;\;\;\;\;\;\;\;\;\;\;\;\;\;\;\;\;\;\;\;\;\;\;\;\;-e^{i(p^{(1)}_F-p^{(2)}_F)(x_2-x_3)/\hbar} \mathcal{Q}_2(x_1, x_2, x_3, x_1-x_2+x_3, t)\Bigg].\label{SM1}
\end{align}
Here the functions $\mathcal{Q}_j$, where $j=1,2$, are defined by
\begin{align}
\mathcal{Q}_1(x_1, x_2, x_3, x_4, t) =&\Big<e^{-i\varphi^\dagger_1(x_1, t)}e^{-i\varphi_1(x_1, t)}\;\;e^{i\varphi^\dagger_2(x_2, t)}e^{i\varphi_2(x_2, t)}\;\;e^{-i\varphi^\dagger_2(x_3, 0)}e^{-i\varphi_2(x_3, 0)}e^{i\varphi^\dagger_1(x_4, 0)}e^{i\varphi_1(x_4, 0)}\Big>,\\
\mathcal{Q}_2(x_1, x_2, x_3, x_4, t){=}& \Big<\!e^{-i\varphi^\dagger_2(x_1, -t)}e^{-i\varphi_2(x_1, -t)}e^{i\varphi^\dagger_1(x_2, -t)}e^{i\varphi_1(x_2, -t)}e^{-i\varphi^\dagger_1(x_3, 0)}e^{-i\varphi_1(x_3, 0)}e^{i\varphi^\dagger_2(x_4, 0)}e^{i\varphi_2(x_4, 0)}\!\Big>.\label{SM2}
\end{align}
Following the standard bosonization approach based on Eqs.~(5) and~(6), we express the functions $\mathcal{Q}_{1,2}$ in terms of the commutators of bosonic fields $\varphi_{ j}$. The expression for $\mathcal{Q}_1$ takes the form
\begin{align}
\mathcal{Q}_1&=e^{\Big[\varphi_2(x_3, 0),\varphi^\dagger_1(x_4, 0)\Big]}e^{\Big[\varphi_1(x_1, t), \varphi^\dagger_2(x_2, t)\Big]}e^{\Big[\varphi_2(x_2, t), \varphi^\dagger_2(x_3, 0)\Big]} e^{-\Big[\varphi_2(x_2, t), \varphi^\dagger_1(x_4, 0)\Big]}e^{-\Big[\varphi_1(x_1, t), \varphi^\dagger_2(x_3, 0)\Big]}e^{\Big[\varphi_1(x_1, t), \varphi^\dagger_1(x_4, 0)\Big]}.\label{SM3}
\end{align}
To account for the interactions both within and between the two systems, we now express the bosonic fields $\varphi_j$ in terms of dressed-bosonic variables $\tilde{b}_{j, l}$ using Eqs.~(6) and~(10). The result can be written as
\begin{align}
\varphi_1(x, t) &=-i\sum_l\frac{\mathcal{C}_l}{\sqrt{l}}e^{\frac{i}{\hbar}\left[q_l x-\tilde{\varepsilon}_l^{(1)}(q_l)t\right]}\tilde{b}_{1, l}+i\sum_l\frac{\mathcal{S}_l}{\sqrt{l}}e^{\frac{i}{\hbar}\left[q_l x-\tilde{\varepsilon}_l^{(2)}(q_l)t\right]}\tilde{b}_{2, l},\label{SM0A}\\
\varphi_2(x, t) &=-i\sum_l\frac{\mathcal{S}_l}{\sqrt{l}}e^{\frac{i}{\hbar}\left[q_l x-\tilde{\varepsilon}_l^{(1)}(q_l)t\right]}\tilde{b}_{1, l}-i\sum_l\frac{\mathcal{C}_l}{\sqrt{l}}e^{\frac{i}{\hbar}\left[q_l x-\tilde{\varepsilon}_l^{(2)}(q_l)t\right]}\tilde{b}_{2, l}.\label{SM0b}
\end{align}
Here the energies $\tilde{\varepsilon}^{(j)}_{ l}$ are defined in Eq.~(11) and we used the shorthand notations $\mathcal{C}_l\equiv \cos\theta_l$ and $\mathcal{S}_l\equiv\sin\theta_l$, where $\theta_l$ is given by Eq.~(10). Using the standard bosonic commutation relations, it is straightforward to show that the equal time commutators in Eq.~\eqref{SM3} vanish, i.e., $\Big[\varphi_2(x_3, 0),\varphi^\dagger_1(x_4, 0)\Big]=0$ and $\Big[\varphi_1(x_1, t), \varphi^\dagger_2(x_2, t)\Big]=0$. The remaining four commutators in Eq.~\eqref{SM3} upon using Eqs.~\eqref{SM0A} and~\eqref{SM0b} become
\begin{align}
\Big[\varphi_2(x_2, t), \varphi^\dagger_2(x_3, 0)\Big] &= \sum_l\Bigg(\frac{\mathcal{S}^2_l}{l} e^{\frac{i}{\hbar}\Big[ q_l(x_2-x_3)-\tilde{\varepsilon}^{(1)}_{ l}t\Big]}+\frac{\mathcal{C}^2_l}{l}e^{\frac{i}{\hbar}\Big[ q_l(x_2-x_3)-\tilde{\varepsilon}^{(2)}_{ l}t\Big]}\Bigg),\\
\Big[\varphi_1(x_1, t), \varphi^\dagger_1(x_4, 0)\Big] &=\sum_{l}\Bigg(\frac{\mathcal{C}^2_l}{l}  e^{\frac{i}{\hbar}\Big[ q_l(x_1-x_4)-\tilde{\varepsilon}^{(1)}_{ l}t\Big]}+\frac{\mathcal{S}^2_l}{l} e^{\frac{i}{\hbar}\Big[ q_l(x_1-x_4)-\tilde{\varepsilon}^{(2)}_{ l}t\Big]}\Bigg),\\
\Big[\varphi_2(x_2, t), \varphi^\dagger_1(x_4, 0)\Big]
&=\sum_l\frac{\mathcal{C}_l\mathcal{S}_l}{l}\Bigg( e^{\frac{i}{\hbar}\Big[ q_l(x_2-x_4)-\tilde{\varepsilon}^{(1)}_{ l}t\Big]}-  e^{\frac{i}{\hbar}\Big[ q_l(x_2-x_4)-\tilde{\varepsilon}^{(2)}_{ l}t\Big]}\Bigg),\\
\Big[\varphi_1(x_1, t), \varphi^\dagger_2(x_3, 0)\Big] 
&=\sum_l\frac{\mathcal{C}_l\mathcal{S}_l}{l}\Bigg( e^{\frac{i}{\hbar}\Big[ q_l(x_1-x_3)-\tilde{\varepsilon}^{(1)}_{ l}t\Big]}-  e^{\frac{i}{\hbar}\Big[ q_l(x_1-x_3)-\tilde{\varepsilon}^{(2)}_{l}t\Big]}\Bigg).\label{SM5}
\end{align}
 The function $\mathcal{Q}_1$ then takes the form
\begin{align}
\mathcal{Q}_1=&\exp \Bigg\{\sum_l\Bigg[\frac{\mathcal{S}^2_l}{l} e^{\frac{i}{\hbar}\Big[ q_l(x_2-x_3)-\tilde{\varepsilon}^{(1)}_{ l}t\Big]}+\frac{\mathcal{C}^2_l}{l}e^{\frac{i}{\hbar}\Big[ q_l(x_2-x_3)-\tilde{\varepsilon}^{(2)}_{ l}t\Big]}+\frac{\mathcal{C}^2_l}{l}  e^{\frac{i}{\hbar}\Big[ q_l(x_1-x_4)-\tilde{\varepsilon}^{(1)}_{ l}t\Big]}+\frac{\mathcal{S}^2_l}{l} e^{\frac{i}{\hbar}\Big[ q_l(x_1-x_4)-\tilde{\varepsilon}^{(2)}_{ l}t\Big]}\nonumber\\
&-\frac{\mathcal{C}_l\mathcal{S}_l}{l}\Bigg( e^{\frac{i}{\hbar}\Big[ q_l(x_2-x_4)-\tilde{\varepsilon}^{(1)}_{ l}t\Big]}-  e^{\frac{i}{\hbar}\Big[ q_l(x_2-x_4)-\tilde{\varepsilon}^{(2)}_{ l}t\Big]}+e^{\frac{i}{\hbar}\Big[ q_l(x_1-x_3)-\tilde{\varepsilon}^{(1)}_{ l}t\Big]}-  e^{\frac{i}{\hbar}\Big[ q_l(x_1-x_3)-\tilde{\varepsilon}^{(2)}_{ l}t\Big]}\Bigg)\Bigg]\Bigg\}.\label{SM6}
\end{align}
Therefore, the required expression of $\mathcal{Q}_1(x_1, x_2, x_3, x_1-x_2+x_3,t)$ for the evaluation of current becomes
\begin{align}
\mathcal{Q}_1(x_1, x_2, x_3, x_1-x_2+x_3,t) &=\exp\Bigg[ \sum_l\frac{1}{l} \Bigg(e^{\frac{i}{\hbar}\Big[ q_l(x_2-x_3)-\tilde{\varepsilon}^{(1)}_{ l}t\Big]}+  e^{\frac{i}{\hbar}\Big[ q_l(x_2-x_3)-\tilde{\varepsilon}^{(2)}_{ l}t\Big]}\Bigg)\nonumber\\
&\;\;\;\;-\sum_l\frac{\mathcal{C}_l\mathcal{S}_l}{l} \Bigg(e^{\frac{i}{\hbar}\Big[ q_l\{2(x_2-x_3)-(x_1-x_3)\}-\tilde{\varepsilon}^{(1)}_{ l}t\Big]}-  e^{\frac{i}{\hbar}\Big[ q_l\{2(x_2-x_3)-(x_1-x_3)\}-\tilde{\varepsilon}^{(2)}_{ l}t\Big]}\Bigg)\nonumber\\
&\;\;\;\;-\sum_l\frac{\mathcal{C}_l\mathcal{S}_l}{l} \Bigg(e^{\frac{i}{\hbar}\Big[ q_l(x_1-x_3)-\tilde{\varepsilon}^{(1)}_{ l}t\Big]}-  e^{\frac{i}{\hbar}\Big[ q_l(x_1-x_3)-\tilde{\varepsilon}^{(2)}_{ l}t\Big]}\Bigg)\Bigg]\nonumber\\
&\equiv \mathcal{Q}_1(x_2-x_3, x_1-x_3, t).\label{SM7}
\end{align}
Following the similar steps outlined above, one can evaluate the expression for $\mathcal{Q}_2(x_1, x_2, x_3, x_1-x_2+x_3, t)$. Finally, we introduce new variables $y=x_2-x_3$ and $y'=x_1-x_3$ to express current per unit length~\eqref{SM1} to the form
\begin{align}
\mathcal{I}
&=\lim_{L\to\infty}\frac{e|\gamma|^2}{  L^3\hbar^2}\int^\infty_{-\infty}dt\; e^{i\frac{\Delta\mu}{\hbar}t}\;\int^{L}_{0}dy\int^{L}_{0}dy' \Big[e^{-i(p^{(1)}_{F}-p^{(2)}_{F})y/\hbar} \mathcal{Q}_1(y, y',t)-e^{i(p^{(1)}_{F}-p^{(2)}_{F})y/\hbar}\mathcal{Q}_1(y, y', -t)\Big].\label{SM9}
\end{align}
In the following, we consider the case of $\Delta\mu=eV>0$. In this case, only the first part of Eq.~\eqref{SM9} contributes to the current. Therefore we proceed with
\begin{align}
\mathcal{I}
=&\lim_{L\to\infty}\frac{e|\gamma|^2}{  L^3\hbar^2}\int^\infty_{-\infty}dt\; e^{i\frac{eV}{\hbar}t}\;\int^{L}_{0}dy\int^{L}_{0}dy' \;e^{-i(p^{(1)}_{F}-p^{(2)}_{F})y/\hbar}\exp \Bigg[\sum_l\frac{1}{l}\Bigg( e^{\frac{i}{\hbar}\Big[ q_l y-\tilde{\varepsilon}^{(1)}_{ l}t\Big]}+  e^{\frac{i}{\hbar}\Big[ q_l  y-\tilde{\varepsilon}^{(2)}_{ l}t\Big]}\Bigg)\Bigg]\nonumber\\
&\times\exp\Bigg[\!\!-\sum_l\frac{\mathcal{C}_l\mathcal{S}_l}{l}\Bigg( e^{\frac{i}{\hbar}\Big[ q_l(2  y- y')-\tilde{\varepsilon}^{(1)}_{ l}t\Big]}-  e^{\frac{i}{\hbar}\Big[ q_l(2  y- y')-\tilde{\varepsilon}^{(2)}_{ l}t\Big]}+ e^{\frac{i}{\hbar}\Big[ q_l y'-\tilde{\varepsilon}^{(1)}_{ l}t\Big]}-  e^{\frac{i}{\hbar}\Big[ q_l y'-\tilde{\varepsilon}^{(2)}_{ l}t\Big]}\Bigg)\Bigg].\label{SM10}
\end{align}
In order to further simplify the Eq.~\eqref{SM10}, we introduce a dimensionless parameter $z=2\pi y'/L$ and express Eq.~\eqref{SM10} in the form
\begin{align}
\mathcal{I}=\lim_{L\to\infty}\frac{e|\gamma|^2}{  2\pi\hbar^2}\frac{1}{L^2}\int^\infty_{-\infty}dt\; &e^{i\frac{eV}{\hbar}t}\;\int^{L}_{0}dy\;e^{-i\Delta p_F y/\hbar}\exp \Bigg[\sum_l\frac{1}{l}\Bigg( e^{i\left[ q_ly-\tilde{\varepsilon}^{(1)}(q_l)t\right]/\hbar}+  e^{i\left[ q
_l y-\tilde{\varepsilon}^{(2)}(q_l)t\right]/\hbar}\Bigg)\Bigg]\nonumber\\
\times\int^{\pi}_{-\pi}dz \;
&\exp\Big[-B(y, z, t; L)\Big],\label{SM11}
\end{align}
where we introduced the function $B(y, z, t; L)$ defined by
\begin{align}
B(y, z, t; L)&= \sum_l\frac{\mathcal{C}_l\mathcal{S}_l}{l}\Big(e^{i\left[q_l(2  y- \frac{L}{2\pi}z)-\tilde{\varepsilon}^{(1)}(q_l)t\right]/\hbar}-  e^{i\left[ q_l(2  y- \frac{L}{2\pi}z)-\tilde{\varepsilon}^{(2)}(q_l)t\right]/\hbar}+ e^{i\left[q_l \frac{L}{2\pi}z-\tilde{\varepsilon}^{(1)}(q_l)t\right]/\hbar}-  e^{i\left[ q_l \frac{L}{2\pi}z-\tilde{\varepsilon}^{(2)}(q_l)t\right]/\hbar}\Big)\nonumber\\
&=\sum_l\frac{\mathcal{C}_l\mathcal{S}_l}{l} \Big(e^{i l\left(\frac{4\pi y}{L}-z\right)}+e^{ilz}\Big)\Big(e^{-i\tilde{\varepsilon}^{(1)}(q_l) t/\hbar}-e^{-i \tilde{\varepsilon}^{(2)}(q_l) t/\hbar}\Big).\label{SM11a}
\end{align}
We note that the variables $t$ and $y$ in Eq.~\eqref{SM11} take the typical values $t\sim\hbar/eV$ and $y\sim\hbar/\Delta p_F$ respectively and thus, do not scale with the system size at $L\to\infty$. The factors $e^{\pm ilz}$ ensure that the sum over $l$ converges at $l\sim 1/|z|$. Thus at $L\to\infty$, we have $q_l\to 0$. In this case, $\mathcal{C}_l$ and $\mathcal{S}_l$ in Eq.~\eqref{SM11a} can be approximated by their respective values in the limit of $q_l\to 0$. The expression inside the second bracket of Eq.~\eqref{SM11a} upon expansion in series in small parameter $q_l$ yields the leading contribution which scales as $L^{-1}$. Therefore in thermodynamic limit, $B(y, z, t; L)$ vanishes and the integral in the second line of Eq.~\eqref{SM11} becomes $2\pi$. The expression for the current per unit length then simplifies to the form
\begin{align}
\mathcal{I}&=\lim_{L\to\infty}\frac{e|\gamma|^2}{\hbar^2}\frac{1}{L^2}\int^\infty_{-\infty}dt\; e^{i\frac{eV}{\hbar}t}\;\int^{L}_{0}dy\;e^{-i\Delta p_F y/\hbar}\exp \Bigg[\sum_l\frac{1}{l}\Big( e^{i\left[ q_l y-\tilde{\varepsilon}^{(1)}(q_l)t\right]/\hbar}+  e^{i\left[q_l y-\tilde{\varepsilon}^{(2)}(q_l)t\right]/\hbar}\Big)\Bigg].\label{SM12}
\end{align}
Equation~\eqref{SM12} upon using the definition of the function $\tilde{A}^{(j)}$ defined in Eq.~(14) yields the required expression for the average current per unit length as given by Eq.~(13) of the main text.
\section{Short range interactions}
\begin{figure}[h]
\begin{center}
\includegraphics[scale=0.8]{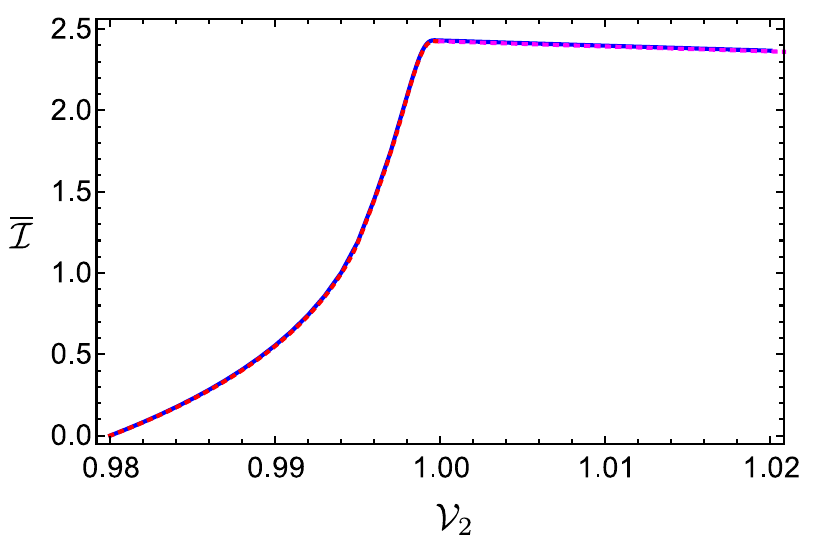}
\caption{The blue line shows the current $\overline{\mathcal{I}}$ as a function of $\mathcal{V}_2$ obtained by performing the integration in  Eq.~\eqref{SMA1} numerically. For this plot we used $\mathcal{V}_1=0.6$, $\mathscr{M}_1=0.01$ and $\mathscr{M}_2=0.02$. For the same parameters, the dashed red and dashed magenta lines show the plot of current given by Eqs.~\eqref{SM27a} and~\eqref{SM22} respectively.}\label{currapp}
\end{center}
\end{figure}

In the main text, we considered the bosonic spectrum for the short range interaction in the form
\begin{equation}
\tilde{\varepsilon}^{(j)}(q)=\tilde{v}_j q-\frac{\tilde{\eta}_j}{2\pi\hbar}q^3,
\end{equation}
where $\tilde{v}_j$ is the function of $v_j$ and $v_{12}$ as given by Eq.~(15) and the parameters $\tilde{\eta}_j$ which account for the non-linearity of bosonic spectrum are given by
\begin{align}
\tilde{\eta}_1&=\frac{1}{2} \left(\eta _1+\eta _2+\frac{\left(\eta _1-\eta _2\right) \left(v_1-v_2\right)+4 \eta _{12} v_{12}}{\sqrt{\left(v_1-v_2\right){}^2+4 v_{12}^2}}\right),\\
\tilde{\eta}_2&=\frac{1}{2} \left(\eta _1+\eta _2-\frac{\left(\eta _1-\eta _2\right) \left(v_1-v_2\right)+4 \eta _{12} v_{12}}{\sqrt{\left(v_1-v_2\right){}^2+4 v_{12}^2}}\right).
\end{align}
Using Eqs.~(13) and~(19), we cast the current per unit length to the form
\begin{align}
\overline{\mathcal{I}} &=\frac{\mathcal{V}_2}{\mathscr{M}_1\mathscr{M}_2}\int^1_0\;dE\int^1_0\;\frac{dQ}{Q^3(1-Q)^3}\;a\left(1{+}\frac{\mathcal{V}_1 E{-}Q}{\mathscr{M}_1 Q^3}\right)a\left(1{+}\frac{\mathcal{V}_2(1{-} E){-}(1{-}Q)}{\mathscr{M}_2 (1-Q)^3}\right),\label{SMA1}
\end{align}
where we introduced the dimensionless parameters
\begin{equation}
\mathscr{M}_{j}{=}\frac{\tilde{\eta}_j}{\tilde{v}_j}\frac{( \Delta p_F)^2}{2\pi\hbar},\mathcal{V}_j{=}\frac{eV}{\tilde{v}_j\Delta p_F},\;\;\overline{\mathcal{I}}=\mathcal{I}\frac{\hbar^2\tilde{v}_1}{e|\gamma|^2}.\label{SMA2}
\end{equation}
The four parameters $\mathcal{V}_j$ and $\mathscr{M}_j$ in the expression of current~\eqref{SMA1} indeed depend on the applied voltage. However, understanding of current~\eqref{SMA1} can also be achieved by considering $\mathcal{V}_j$ and $\mathscr{M}_j$ as free parameters. The behavior of current $\overline{\mathcal{I}}$ as a function of $\mathcal{V}_2$ for fixed $\mathcal{V}_1$ and $\mathscr{M}_j$ is illustrated in Fig.~\ref{currapp}.

\begin{figure}[t]
\begin{center}
\includegraphics[scale=1.2]{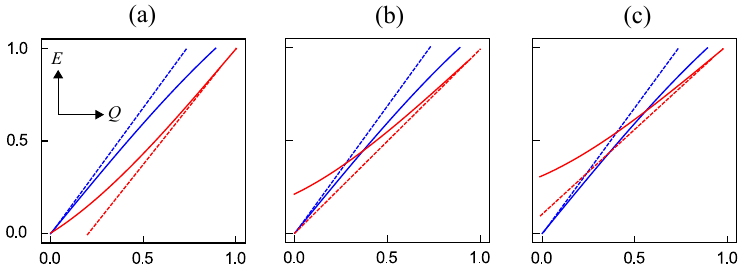}
\caption{Plots of equations $\mathcal{V}_1E-Q+\mathscr{M}_1Q^3=0$ (solid blue) and $\mathcal{V}_2(1-E)-(1-Q)+\mathscr{M}_2(1-Q)^3=0$ (solid red) for $\mathcal{V}_1<\mathcal{V}_2$ and $\mathscr{M}_1<\mathscr{M}_2$. Corresponding asymptotes in the limit of $\mathscr{M}_{1, 2}=0$ are represented by dashed lines. In the plots, we assumed $\mathcal{V}_1$ and $\mathscr{M}_1$ are fixed and (a) $\mathcal{V}_2=1-\mathscr{M}_2$, (b) $\mathcal{V}_2=1$ and (c) $\mathcal{V}_2=1+\mathscr{M}_2$.}\label{supfig1}
\end{center}
\end{figure}

In the following, we provide the quantitative understanding for the behavior of current shown in Fig.~\ref{currapp}. To this end, we first note that the function $a(x)$ in the expression of current~\eqref{SMA1} vanishes outside the interval $0<x<1$. Therefore, to get finite current, we are interested in the regime where the argument of both $a\left(1{+}\frac{\mathcal{V}_1 E{-}Q}{\mathscr{M}_1 Q^3}\right)$ and $a\left(1{+}\frac{\mathcal{V}_2(1{-} E){-}(1{-}Q)}{\mathscr{M}_2 (1-Q)^3}\right)$ lines within $0$ and $1$. It is then clear that for the current~\eqref{SMA1} to be finite, there should exists an overlapping region defined by the four equalities $1{+}\frac{\mathcal{V}_1 E{-}Q}{\mathscr{M}_1 Q^3}=0, 1$ and $1{+}\frac{\mathcal{V}_2(1{-} E){-}(1{-}Q)}{\mathscr{M}_2 (1-Q)^3}=0, 1$. These four equations are plotted in Fig.~\ref{supfig1} for different choice of the parameter $\mathcal{V}_2$ keeping $\mathcal{V}_1$ as fixed. From this figure, it is easy to see that the minimum value of $\mathcal{V}_2$ for the current to be finite corresponds to the case where the solid red line passes through the origin. This minimum value of $\mathcal{V}_2$ can be evaluated straightforwardly
\begin{equation}
\left(1+\frac{\mathcal{V}_2(1- E)-(1-Q)}{\mathscr{M}_2 (1-Q)^3}\right)\Bigg|_{Q=0, E=0}=0\implies \mathcal{V}_2=1-\mathscr{M}_2.\label{kk2}
\end{equation}
Increasing $\mathcal{V}_2$ from its threshold value given by Eq.~\eqref{kk2}, current starts to grow in a rather non-trivial fashion due to the non-linearity of bosonic spectrum.
Upon further increasing $\mathcal{V}_2$ such that $\mathcal{V}_2\geq 1$, the area of overlapping region between four lines reaches approximately to the constant value as illustrated in Figs.~\ref{supfig1}(b) and (c). This implies that the current acquires its plateau behavior. To gain more insights on the behavior of current~\eqref{SMA1}, in the following, we study these two cases (i) $\mathcal{V}_2\leq 1$ and (ii) $\mathcal{V}_2\geq 1$ separately.
\subsection{The case of $\mathcal{V}_2\leq 1$}
In this case, as discussed in the main text and also apparemt from Fig.~\ref{supfig1}(a), $a(1{+}\frac{\mathcal{V}_1 E{-}Q}{\mathscr{M}_1 Q^3})$ in Eq.~\eqref{SMA1} can be replaced by the delta function peaked at $E=Q/\mathcal{V}_1$, i.e, $\mathscr{M}_1\to 0$. Thus the expression of the current~\eqref{SMA1} simplifies to
\begin{align}
\overline{\mathcal{I}} &=\frac{\mathcal{V}_2}{\mathcal{V}_1}\int^1_0\;\frac{dQ}{\mathscr{M}_2(1-Q)^3}\;a\left(\frac{\mathcal{V}_2-1-Q\left( \frac{\mathcal{V}_2}{\mathcal{V}_1}-1\right)}{\mathscr{M}_2 (1-Q)^3}+1\right).\label{SM23}
\end{align}
To make further progress, we proceed with the substitution
\begin{equation}
x= \frac{Q\left( \frac{\mathcal{V}_2}{\mathcal{V}_1}-1\right)-(\mathcal{V}_2-1)}{\mathscr{M}_2 (1-Q)^3},\label{SM23a}
\end{equation}
which transforms Eq.~\eqref{SM23} to the form
\begin{equation}\label{SM24}
\overline{\mathcal{I}}  =\mathcal{V}_2 \int^1_{\frac{1-\mathcal{V}_2}{\mathscr{M}_2}} dx\;a(1-x)\mathcal{K}_1(x).
\end{equation}
Here the Jacobian of the transformation~\eqref{SM23a} is denoted by the function $\mathcal{K}_1(x)$ and is given by
\begin{equation}
\mathcal{K}_1(x)=\Bigg|\frac{1-Q(x)}{\mathcal{V}_1(2-3\mathcal{V}_2)+\mathcal{V}_2+2Q(x)(\mathcal{V}_2-\mathcal{V}_1)}\Bigg|,\quad Q(x)=\frac{\mathcal{V}_2-1+x\mathscr{M}_2(1-Q)^3}{\frac{\mathcal{V}_2}{\mathcal{V}_1}-1}.\label{SM25}
\end{equation}
We then treat the function $\mathcal{K}_1$ perturbatively in small parameter $\mathscr{M}_2$, which gives up on successive iterations
\begin{align}
Q_0 &=\frac{\mathcal{V}_2-1}{\frac{\mathcal{V}_2}{\mathcal{V}_1}-1},\nonumber\\
Q_1 &=Q_0+\frac{x \mathscr{M}_2 (1-Q_0)^3}{\frac{\mathcal{V}_2}{\mathcal{V}_1}-1},\nonumber\\
Q_2 &=Q_0+\frac{x \mathscr{M}_2 (1-Q_1)^3}{\frac{\mathcal{V}_2}{\mathcal{V}_1}-1}.\label{SM26}
\end{align}
Using Eqs.~\eqref{SM26} and~\eqref{SM25}, we get the expression of $\mathcal{K}_1$ to the first order in $\mathscr{M}_2$, which is given by
\begin{align}
\mathcal{K}_1(x)
&=-\frac{\mathscr{M}_2\left(1-\mathcal{V}_1\right)^2\mathcal{V}_1(\mathcal{V}_2)^2 x-\left(\mathcal{V}_2-\mathcal{V}_1\right)^3}{\left(\mathcal{V}_2-\mathcal{V}_1\right)^4+2\mathscr{M}_2 x\left(1-\mathcal{V}_1\right)^2\mathcal{V}_1(\mathcal{V}_2)^2
\left(\mathcal{V}_2-\mathcal{V}_1\right)}\simeq \frac{1}{\mathcal{V}_2-\mathcal{V}_1}-\frac{3\mathscr{M}_2(1-\mathcal{V}_1)^2\mathcal{V}_1(\mathcal{V}_2)^2}{\left(\mathcal{V}_2-\mathcal{V}_1\right)^4}x.\label{SM27}
\end{align}
Substitution of Eq.~\eqref{SM27} into the expression current~\eqref{SM24} yields
\begin{equation}\label{SM27a}
\overline{\mathcal{I}} = \frac{\mathcal{V}_2}{\mathcal{V}_2-\mathcal{V}_1}\int^1_{\frac{1-\mathcal{V}_2}{\mathscr{M}_2}} dx\;a(1-x)-\frac{3\mathscr{M}_2(1-\mathcal{V}_1)^2\mathcal{V}_1\mathcal{V}_2^3}{\left(\mathcal{V}_2-\mathcal{V}_1\right)^4}\int^1_{\frac{1-\mathcal{V}_2}{\mathscr{M}_2}} dx\;a(1-x)\;x,
\end{equation}
For $\mathcal{V}_2$ close to $1$, the second term of above equation, which is linear in $\mathscr{M}_2$, can be neglected. The remaining first term upon introducing $\lambda$ such that
\begin{equation}
\mathcal{V}_2=1+\lambda \mathscr{M}_2,
\end{equation}
reproduces Eq.~(20) in the main text. Therefore, when $\mathcal{V}_2$ is close to $1$, current~\eqref{SMA1} can be approximated by
\begin{equation}\label{SM28}
\overline{\mathcal{I}} = \frac{1}{1-\mathcal{V}_1}\int^{1+\lambda}_{0}dx\;a(x).
\end{equation}
As shown in Fig.~\ref{currapp}, the behavior current as a function of $\mathcal{V}_2\leq 1$ given by Eq.~\eqref{SM27a} coincides with that obtained by performing the integration in  Eq.~\eqref{SMA1} numerically. For $\mathcal{V}_2\geq 1$, as demonstrated in Figs.~\ref{currapp}(b) and (c), both parameters $\mathscr{M}_j$ should be treated in equal footing.
\subsection{ The case of $\mathcal{V}_2\geq 1$}
Here we treat the case of both $\mathscr{M}_j$ being finite to get current~\eqref{SMA1} in the limit of $\mathcal{V}_2\geq 1$. To this end, we first perform the coordinate transformations
\begin{align}
\frac{Q-\mathcal{V}_1 E}{\mathscr{M}_1 Q^3}=x,\;\;\;\frac{(1-Q)-\mathcal{V}_2(1- E)}{\mathscr{M}_2 (1-Q)^3}=y,\label{SM15}
\end{align}
which gives
\begin{equation}
Q(x, y)=\frac{\mathcal{V}_1(\mathcal{V}_2-1)+\mathscr{M}_1 \mathcal{V}_2 Q^3 x+\mathscr{M}_2 \mathcal{V}_1(1-Q)^3y}{\mathcal{V}_2-\mathcal{V}_1},\;\;\;E(x, y)=\frac{Q-\mathscr{M}_1 Q^3x}{\mathcal{V}_1}.\label{SM16}
\end{equation}
Under the transformations~\eqref{SM15}, the expression of the current~\eqref{SMA1} becomes
\begin{align}\label{SM17}
\overline{\mathcal{I}}&=\mathcal{V}_2\int^1_0\;dx\int^1_0\;dy\;a(1-x)\;a(1-y)\mathcal{K}_2(x, y),
\end{align}
where the Jacobian of transformation is given by
\begin{equation}
\mathcal{K}_2(x, y)=\Bigg|\frac{Q(1-Q)}{Q\left[\mathcal{V}_1(2-3\mathcal{V}_2)-2\mathcal{V}_2\right]-2Q^2(\mathcal{V}_1-\mathcal{V}_2)+
3\mathcal{V}_1\mathcal{V}_2 E}\Bigg|.\label{SM18}
\end{equation}
We evaluate $\mathcal{K}_2$ perturbatively in the small parameters $\mathscr{M}_{j}$. Using Eqs.~\eqref{SM16} and~\eqref{SM18}, we obtain the following iterative solutions for $Q$ and $E$

\begin{align}
Q_0=\frac{\mathcal{V}_1(\mathcal{V}_2-1)}{\mathcal{V}_2-\mathcal{V}_1},\;\;E_0=\frac{\mathcal{V}_2-1}{\mathcal{V}_2-\mathcal{V}_1},
\end{align}

\begin{align}
Q_1=Q_0+\frac{\mathscr{M}_1 \mathcal{V}_2 Q^3_0 x+\mathscr{M}_2 \mathcal{V}_1(1-Q_0)^3 y}{\mathcal{V}_2-\mathcal{V}_1},\;\;\;\;\;E_1=\frac{Q_0-\mathscr{M}_1 Q^3_0 x}{\mathcal{V}_1},\label{SM20}
\end{align}

\begin{equation}
Q_2=Q_0+\frac{\mathscr{M}_1 \mathcal{V}_2 Q^3_0 x+\mathscr{M}_2 \mathcal{V}_1(1-Q_0)^3 y}{\mathcal{V}_2-\mathcal{V}_1},\;\;\;\;E_2=\frac{Q_1-\mathscr{M}_1 Q^3_0 x}{\mathcal{V}_1}.\label{SM20a}
\end{equation}
Using Eqs.~\eqref{SM18} and~\eqref{SM20a}, we obtained the expression of $\mathcal{K}_2$ to the linear order in the parameters $\mathscr{M}_j$, which is given by
\begin{align}
\mathcal{K}_2(x, y)=\frac{1}{\mathcal{V}_2-\mathcal{V}_1}+\frac{3(\mathcal{V}_1)^2\mathcal{V}_2(\mathcal{V}_2-1)^2\;x}{(\mathcal{V}_2-\mathcal{V}_1)^4}\mathscr{M}_1-\frac{3(\mathcal{V}_2)^2\mathcal{V}_1(1-\mathcal{V}_1)^2\;y}{(\mathcal{V}_2-\mathcal{V}_1)^4}\mathscr{M}_2.\label{SM21}
\end{align}
Substitution of Eq.~\eqref{SM21} into Eq.~\eqref{SM17} gives
\begin{equation}\label{SM22}
\overline{\mathcal{I}} =\frac{\mathcal{V}_2}{\mathcal{V}_2-\mathcal{V}_1}+\frac{(\mathcal{V}_1)^2(\mathcal{V}_2)^2(\mathcal{V}_2-1)^2}{(\mathcal{V}_2-\mathcal{V}_1)^4}\mathscr{M}_1-\frac{(\mathcal{V}_2)^3\mathcal{V}_1(1-\mathcal{V}_1)^2}{(\mathcal{V}_2-\mathcal{V}_1)^4}\mathscr{M}_2.
\end{equation}
To arrive at Eq.~\eqref{SM22}, we used the identity $\int^1_0 dx\;a(1-x)x=1/3$.
For $\mathscr{M}_1\to 0$, the correction given by Eq.~\eqref{SM22} and that by Eq.~\eqref{SM27a} are formally identical. We note that, the limit $\mathcal{V}_2\to 1$ in Eq.~\eqref{SM22} is independent of $\mathscr{M}_1$. The corrections given by the second and third terms of Eq.~\eqref{SM22} fully account for the downward tilt of current plateau as illustrated in Fig.~\ref{currapp}.
\section{Voltage dependence of $\Delta p_F$}
In this section we outline the derivation for the voltage dependence of $\Delta p_F$ given by Eq.~(16) for the case of short range interactions. It is important to note that in Eq.~(1) of the main text, we omitted the $q=0$ components since they are of no importance for the calculation of transport properties. The omitted part of Hamiltonian~(1) accounts for the dependence of the energy $\mathscr{E}$ of the system on the number of particles in each of the chiral branches and has the form
\begin{align}
\mathscr{E}= 2\pi\hbar\Bigg[ \frac{ v_{1}}{2L} \left(N_1\right)^2 + \frac{v_{2}}{2L} \left(N_2\right)^2+\frac{v_{12}}{L}N_1N_2\Bigg].\label{SM38}
\end{align}
Here the numbers of particles $N_1$ and $N_2$ are measured from those in the ground state of the system with the chemical potentials $\mu_1=\mu_2=0$. The parameters $v_j$ and $v_{12}$ are defined in Eq.~(15). The chemical potentials $\mu_j$ are
\begin{align}
\mu_1 =\frac{\partial \mathscr{E}}{\partial N_1} &=2\pi\hbar\Bigg[\frac{ v_{1}}{L} N_1+\frac{v_{12}}{L}N_2\Bigg],\label{SM38a}\\
\mu_2 =\frac{\partial \mathscr{E}}{\partial N_2} &=2\pi\hbar\Bigg[\frac{ v_{2}}{L} N_2+\frac{v_{12}}{L}N_1\Bigg].\label{SM38b}
\end{align}
We then set $\mu_2=0$ and define the voltage as $eV=\mu_1-\mu_2$. Equation~\eqref{SM38b} then results in the relation 
\begin{equation}
N_2=-\frac{v_{12}}{ v_{2}}N_1.\label{SM39}
\end{equation}
From Eqs.~\eqref{SM38a}$-$\eqref{SM39}, we obtain
\begin{equation}
\frac{N_1}{L}=\frac{eV}{2\pi\hbar}\Bigg[\frac{v_{2}}{v_{1}v_{2}-\left(v_{12}\right)^2}\Bigg],\quad
\frac{N_2}{L}=-\frac{eV}{2\pi\hbar}\Bigg[\frac{v_{12}}{v_{1}v_{2}-\left(v_{12}\right)^2}\Bigg].\label{SM40}
\end{equation}
The shift of Fermi momentum due to the application of voltage can be obtained from the corresponding particle number as
\begin{equation}
p_F^{(1)}(eV)-p_F^{(1)}(0)=2\pi\hbar \frac{N_1}{L}=eV\Bigg[\frac{v_{2}}{v_{1}v_{2}-\left(v_{12}\right)^2}\Bigg].\label{SMaa1}
\end{equation}
Similarly, we have
\begin{equation}
p_F^{(2)}(eV)-p_F^{(2)}(0)=-eV\Bigg[\frac{v_{12}}{v_{1}v_{2}-\left(v_{12}\right)^2}\Bigg].\label{SMaa2}
\end{equation}
Subtraction of Eq.~\eqref{SMaa2} from Eq.~\eqref{SMaa1} results in Eq.~(16) of the main text.
\end{document}